\documentclass[12pt,thmsa]{article}
\usepackage{sw20aip}
\usepackage{amsmath}
\usepackage{graphicx}
\usepackage{amsfonts}
\usepackage{amssymb}
\begin{document}

\begin{center}
{\Large Evolution of the vorticity-area density during the formation of
coherent structures in two-dimensional flows}

\smallskip by

{\large H.W. Capel}

Institute of Theoretical Physics, Univ. of Amsterdam

Valckenierstraat 65, 1018 XE Amsterdam, The Netherlands

and

{\large R.A. Pasmanter}\footnote{Corresponding author. E-mail: pasmante@knmi.nl}

KNMI, P.O.Box 201, 3730 AE De Bilt, The Netherlands

\medskip Abstract
\end{center}

\noindent It is shown: 1) that in two-dimensional, incompressible, viscous
flows the vorticity-area distribution evolves according to an
advection-diffusion equation with a negative, time dependent diffusion
coefficient and 2) how to use the vorticity-streamfunction relations, i.e.,
the so-called scatter-plots, of the quasi-stationary coherent structures in
order to quantify the experimentally observed changes of the vorticity
distribution moments leading to the formation of these structures.

\begin{center}
\newpage
\end{center}

\section{\medskip Introduction\label{introduction}}

Numerical simulations of the freely decaying, incompressible Navier-Stokes
equations in two dimensions have shown that under appropriate conditions and
after a relatively short period of chaotic mixing, the vorticity becomes
strongly localized in a collection of vortices which move in a background of
weak vorticity gradients [\cite{mcwilliams}]. As long as their sizes are much
smaller than the extension of the domain, the collection of vortices may
evolve self-similarly in time [\cite{santangelo} ,\cite{car1} ,\cite{car2}
,\cite{cardoso} ,\cite{borue} ,\cite{Warn}] until one large-scale structure
remains. If the corresponding Reynolds number is large enough, the time
evolution of these so-called coherent structures is usually given by a uniform
translation or rotation and by relatively slow decay and diffusion, the last
two are due to the presence of a non-vanishing viscosity. In other words, in a
co-translating or co-rotating frame of reference, one has quasi-stationary
structures (QSS) which are, to a good approximation, stationary solutions of
the inviscid Euler equations. Accordingly, their corresponding
vorticity\footnote{We use the subscript $S$ in order to indicate that the
field corresponds to an observed QSS.} fields $\omega_{S}(x,y)$ and stream
functions $\psi_{S}(x,y)$ are, to a good approximation, functionally related,
i.e., $\omega_{S}(x,y)\approx\omega_{S}(\psi_{S}(x,y)).$ Similar phenomena
have been observed in the quasi two-dimensional flows studied in the
laboratory [\cite{flor} ,\cite{marteau}]. The only exception to this rule is
provided by the large-scale, oscillatory states that occasionally result at
the end of the chaotic mixing period [\cite{segre} ,\cite{brands2}]. In many
cases, e.g., when the initial vorticity field is randomly distributed in
space, the formation of the QSS corresponds to the segregation of
different-sign vorticity and the subsequent coalescence of equal-sign
vorticity, i.e., to a \emph{spatial demixing} of vorticity.\newline Besides
the theoretical fluid-dynamics context, a good understanding of the
above-described process has implications in many other physically interesting
situations like: geophysical flows [\cite{hopf}], plasmas in magnetic fields
[\cite{kraich2} ], galaxy structure [\cite{stellar}], etc. For these reasons
numerical and experimental studies are still being performed and have already
led to a number of ``scatter plots'', i.e., to the determination of the
$\omega_{S}$-$\psi_{S}$ functional relation as a characterization of the QSS
which appear under different circumstances. Simultaneously, on the theoretical
side, approaches have been proposed which attempt at, among other things,
predicting the QSS directly from the initial vorticity field; if successful in
this, such methods would also alleviate the need of performing costly
numerical and laboratory studies.

The above-mentioned studies point out the large enstrophy decay that often
takes place during the formation of the QSS; sometimes, also the evolution of
the skewness is reported. But for these two lowest-order moments, little
attention has been paid to the evolution of the vorticity-area distribution,
defined in equation (\ref{G}), during the formation of the QSS. In the context
of 2D flows, this distribution plays a very important role: with appropriate
boundary conditions, it is conserved by the inviscid Euler equations and the
stationary solutions are the maximizers of the energy for the given vorticity
distribution [\cite{Arnold1} ,\cite{Arnold2}], see also Subsection
\ref{vorti-redistribu}\ref{general vorti}.

In the present work we study the time evolution of the vorticity-area
distrbution in two-dimensional, incompressible and viscous fluids. Many of the
ideas we present should be applicable also to more realistic systems, e.g.,
when potential vorticity is the Lagrangian invariant. The paper is structured
as follows: in the following Section, we derive, from the Navier-Stokes
equation, the time evolution of the vorticity-area density. It is an
advection-diffusion equation with a time dependent, \emph{negative} diffusion
coefficient. For the purpose of illustration, explicit calculations are
presented in Subsection \ref{Vorticity-area}\ref{Gaussian} for the case of a
Gaussian monopole and, in Subsection \ref{Vorticity-area}\ref{self-similar},
for the case of self-similar decay described by Bartello and Warn in
[\cite{Warn}]. Based on these diffusion coefficients, it would be interesting
to set up a classification distinguishing different various typical scenarios
leading to the QSS. Considering the QSS, a very natural but difficult question
arises about its relation to theoretical predictions, namely, it is not
trivial to quantify how good or bad the agreement between observation and
prediction is, as already stressed, e.g., in [\cite{Majda} ]. For this and
related reasons, in Section \ref{vorti-redistribu}, we show that a perfect
agreement between an observed QSS and the corresponding prediction obtained
through the statistical-mechanics approach as developed by J. Miller et al.
[\cite{millerPRL} ,\cite{robertJSP} ,\cite{millerPRA}] and by Robert and
Sommeria [\cite{robertfrans} ,\cite{robsom}] would imply the equality of the
difference in the moments of the initial and final vorticity distributions on
the one hand and a set of quantities that can be directly obtained from the
experimental $\omega_{S}$-$\psi_{S}$ relation on the other side. The details
of the proof can be found in the Appendix. In \ref{vorti-redistribu}\ref{new
ExperimentAnalysis}, we discuss how to use these quantities as yardsticks in
order to quantify the validity of the statistical-mechanics approach in
numerical and laboratory experiments. In the last Section we summarize our
results and add some comments.

\section{Vorticity-area distribution\label{Vorticity-area}}

\subsection{Time evolution\label{anti-diffusion}}

It turns out that the vorticity-area density undergoes an anti-diffusion
process as we next show. The time dependent vorticity-area density
$G(\sigma,t)$ is given by
\begin{equation}
G(\sigma,t):=\int_{A}\!dxdy\,\delta(\sigma-\omega(x,y,t)),\label{G}%
\end{equation}
where $A$ denotes the domain and the vorticity field $\omega(x,y,t)\mathbf{:=}%
\vec{k}\cdot(\mathbf{\nabla}\times\vec{v})$ evolves according to the
Navier-Stokes equation
\begin{equation}
\frac{\partial\omega}{\partial t}+\vec{v}\cdot\mathbf{\nabla}\omega=\nu
\Delta\omega,\label{NS}%
\end{equation}
where the incompressibility condition $\mathbf{\nabla\cdot}\vec{v}=0$ has been
taken into account. The Navier-Stokes equation determines the time evolution
of the vorticity-area density; one has
\begin{align*}
\frac{\partial G(\sigma,t)}{\partial t} &  =\int_{A}\!dxdy\,\delta^{\prime
}(\sigma-\omega(x,y,t))\vec{v}\cdot\mathbf{\nabla}\omega-\nu\int
_{A}\!dxdy\,\delta^{\prime}(\sigma-\omega(x,y,t))\Delta\omega\\
&  =-\int_{A}\!dxdy\,\vec{v}\cdot\mathbf{\nabla}\delta(\sigma-\omega
(x,y,t))-\nu\frac{\partial\;}{\partial\sigma}\int_{A}\!dxdy\,\delta
(\sigma-\omega(x,y,t))\Delta\omega.
\end{align*}
We will assume impermeable boundaries $\partial A$, i.e., that the velocity
component perpendicular to $\partial A$ vanishes. Therefore, the first
integral in the last expression is zero. Partial integration of the second
integral leads to
\begin{equation}
\frac{\partial G(\sigma,t)}{\partial t}=-\nu\frac{\partial^{2}\;}%
{\partial\sigma^{2}}\int_{A}\!dxdy\,\delta(\sigma-\omega(x,y,t))\left|
\mathbf{\nabla}\omega\right|  ^{2}-\nu\frac{\partial\;}{\partial\sigma}%
\oint_{\partial A}\delta(\sigma-\omega)\mathbf{\nabla}\omega\cdot\vec{n}\,dl.
\end{equation}
The last term represents the net vorticity generation or destruction that
occurs on the boundary $\partial A$ of the domain, $\vec{n}$ is the outward
oriented, unit vector normal to the boundary. This source term vanishes in
some special cases like doubly periodic boundary conditions or when the
support of the vorticity field remains always away from the boundary.
\newline From the definition of $G(\sigma,t)$ one sees that if $G(\sigma,t)=0$
and $\left|  \mathbf{\nabla}\omega\right|  $ is finite, then also the
integrals in the last expression must vanish. Consequently, we can define an
effective diffusion coefficient $D(\sigma,t)$ through
\begin{equation}
\nu\int_{A}\!dxdy\,\delta(\sigma-\omega(x,y,t))\left|  \mathbf{\nabla}%
\omega\right|  ^{2}=:D(\sigma,t)G(\sigma,t),\label{difcoef}%
\end{equation}
From this definiton it is clear that $D(\sigma,t)\geq0$ is the average of
$\nu\left|  \nabla\omega\right|  ^{2}$ over the area on which the vorticity
takes the value $\sigma.$ Therefore, the time evolution of $G(\sigma,t)$ can
be written as an advection-diffusion equation in $\sigma$-space with a source
term, i.e.,
\begin{equation}
\frac{\partial G(\sigma,t)}{\partial t}=-\frac{\partial\;}{\partial\sigma
}\left[  s(\sigma,t)+\frac{\partial D(\sigma,t)}{\partial\sigma}%
G(\sigma,t)+D(\sigma,t)\frac{\partial G(\sigma,t)}{\partial\sigma}\right]
,\label{dG/dt}%
\end{equation}
with a \emph{negative }diffusion coefficient $-D(\sigma,t),$ a ``velocity
field'' $\partial D(\sigma,t)/\partial\sigma$ and where minus the $\sigma
$-derivative of $s(\sigma,t):=\nu\oint_{\partial A}\delta(\sigma
-\omega)\mathbf{\nabla}\omega\cdot\vec{n}\,dl$ is the vorticity source at the boundary.

Introducing the vorticity moments
\[
\Gamma_{m}(t):=\int\!d\sigma\,\sigma^{m}G(\sigma,t)=\int_{A}\!dxdy\,\omega
^{m}(x,y,t),
\]
one can check that the equations above imply that the first moment of the
distribution $G(\sigma,t),$ i.e., the total circulation $\Gamma_{1}$, evolves
according to
\[
\frac{d\Gamma_{1}}{dt}=\int\!d\sigma\,s(\sigma,t),
\]
and that the even moments $\Gamma_{2n}(t):=\int\!d\sigma\,\sigma^{2n}%
G(\sigma,t)$ change in time according to
\[
\frac{d\Gamma_{2n}}{dt}=2n\int\!d\sigma\,\sigma^{m-1}s(\sigma,t)-\nu
2n(2n-1)\int\!dxdy\,\omega^{2(n-1)}\left|  \mathbf{\nabla}\omega\right|  ^{2}.
\]
In particular, when the boundary source term $s(\sigma,t)$ vanishes, the total
circulation $\Gamma_{1}$ is conserved and the even moments \emph{decay} in
time
\begin{equation}
\frac{d\Gamma_{2n}}{dt}=-\nu2n(2n-1)\int\!dxdy\,\omega^{2(n-1)}\left|
\mathbf{\nabla}\omega\right|  ^{2}\leq0.\label{decay}%
\end{equation}
Therefore, in $\sigma$-space one has anti-diffusion at all times and with
$\left|  \mathbf{\nabla}\omega\right|  \rightarrow_{t\rightarrow\infty}0$ when
the boundary source term $s(\sigma,t)$ vanishes, the final vorticity-area
distribution is
\[
\lim_{t\rightarrow\infty}G(\sigma,t)=A\delta(\sigma-\bar{\omega}%
)\quad\mathrm{and}\quad\lim_{t\rightarrow\infty}D(\sigma,t)=0,
\]
where $A$ is the area of the domain and $\bar{\omega}$ is the average
vorticity, i.e., $\bar{\omega}=\Gamma_{1}/A.$

In most cases, it is not possible to make an a priori calculation of the
effective diffusion coefficient $D(\sigma,t).$ On the other hand, its
computation is straightforward if a solution of the Navier-Stokes equation is
known. These diffusion coefficients and, in particular the product
$D(\sigma,t)G(\sigma,t),$ confer Figs. 1 and 2 in \ref{self-similar}, may lead
to a classification of different scenarios for and stages in the formation of
QSS. It is also worthwhile recalling that conditional averages very similar to
$D(\sigma,t),$ confer (\ref{difcoef}), play an important role in, e.g., the
advection of passive scalars by a random velocity field. In the case of
passive-scalar advection by self-similar, stationary turbulent flows,
Kraichnan has proposed a way of computing such quantities [ \cite{Kraichnan}%
].\newline For the purpose of illustration, in the next Subsection, we use an
exact analytic solution in order to compute the corresponding $G(\sigma,t)$
and $D(\sigma,t).$ Another tractable case occurs when the flow evolves
self-similarly; in Subsection \ref{Vorticity-area}\ref{self-similar} we apply
these ideas to the self-similar evolution data obtained by Bartello and Warn
[\cite{Warn}].

\subsection{A simple example\label{Gaussian}}

As a simple example consider the exact solution of the Navier-Stokes equation
in an infinite domain given by a Gaussian monopole with circulation
$\Gamma_{1}$,
\[
\omega_{G}(x,y,t)=\Gamma_{1}\left(  4\pi\nu t\right)  ^{-1}\exp(-r^{2}/4\nu
t),\;\mathrm{with}\;r^{2}:=x^{2}+y^{2}\;\mathrm{and}\;t\geq0.
\]
Then, in cylindrical coordinates $(r,\phi),$%
\begin{align*}
\delta(\sigma-\omega_{G}(x,y,t))\,dxdy &  =\delta(r^{2}-R^{2}(\sigma
,t))\,\frac{1}{2}dr^{2}d\phi/\left|  \frac{\partial\omega_{G}}{\partial r^{2}%
}\right|  _{r^{2}=R^{2}},\\
\mathrm{where}\;\left.  \omega_{G}(x,y,t)\right|  _{r=R} &  =\sigma
,\;\mathrm{i.e.,}\;R^{2}(\sigma,t):=-4\nu t\ln(\frac{\sigma4\pi\nu t}%
{\Gamma_{1}})\\
\mathrm{and}\;\left.  \frac{\partial\omega_{G}}{\partial r^{2}}\right|
_{r^{2}=R^{2}} &  =-\frac{\sigma}{4\nu t},
\end{align*}
and we have that for such a Gaussian monopole, the vorticity-area density is
\begin{align*}
G_{G}(\sigma,t) &  =0,\;\mathrm{for}\;\sigma<0,\\
&  =4\pi\nu t\sigma^{-1},\;\mathrm{for}\;0<\sigma\leq\sigma_{\max}(t),\\
&  =0,\;\mathrm{for}\;\sigma_{\max}(t)<\sigma,\\
\mathrm{with}\;\sigma_{\max}(t) &  \equiv\Gamma_{1}\left(  4\pi\nu t\right)
^{-1}.
\end{align*}
The divergence at $\sigma\rightarrow0$ is due to the increasingly large areas
occupied by vanishingly small vorticity associated with the tails of the
Gaussian profile. Due to this divergence, the density $G_{G}(\sigma,t)$ is not
integrable as it should since the domain $A$ is infinite. In spite of this
divergence, all the $\sigma$-moments are finite, in particular, the first
moment, i.e., the circulation, equals $\Gamma_{1}$ and the second moment,
i.e., the enstrophy, is $\Gamma_{1}^{2}/8\pi\nu t.$ As expected, the
circulation is constant in time while the enstrophy decays to zero\footnote{By
contrast, the \emph{spatial} second moments of $\omega(x,y,t)$ \emph{increase}
in time like $2\nu t$, e.g., $\int\int\!dxdyx^{2}\omega_{G}(x,y,t)=2\nu t.$}.
It is also interesting to notice that while the maximum vorticity value,
$\sigma_{\max}(t)=\Gamma_{1}/4\pi\nu t,$ occupies only one point, i.e., a set
of zero dimension, the density $G_{G}(\sigma,t)$ remains finite for
$\sigma\nearrow\sigma_{\max}(t),$ more precisely, $\lim_{\sigma\nearrow
\sigma_{\max}(t)}G_{G}(\sigma,t)=\left(  4\pi\nu t\right)  ^{2}/\Gamma_{1}.$

Moreover, in this simple example one can also compute
\[
\nu\int\!dxdy\,\delta(\sigma-\omega_{G}(x,y,t))\left|  \mathbf{\nabla}%
\omega_{G}\right|  ^{2}=4\pi\nu\sigma\ln\left(  \frac{\sigma_{\max}(t)}%
{\sigma}\right)  ,\;\mathrm{for}\;0<\sigma\leq\sigma_{\max}(t),
\]
so that the corresponding effective diffusion coefficient is
\[
D_{G}(\sigma,t)=\frac{\sigma^{2}}{t}\ln\left(  \frac{\sigma_{\max}(t)}{\sigma
}\right)  ,\;\mathrm{for}\;0<\sigma\leq\sigma_{\max}(t).
\]
The vanishing of this $D_{G}(\sigma,t)$ with $\sigma\rightarrow0$ corresponds
to the vanishingly small spatial gradients of vorticity at large distances
from the vortex core; this gradient vanishes also at the center of the vortex
leading to a (weaker) vanishing of $D_{G}(\sigma,t)$ for $\sigma\nearrow
\sigma_{\max}(t)=\Gamma_{1}/4\pi\nu t.$ The Gaussian vortex is totally
dominated by viscosity and yet the corresponding effective diffusion
coefficient $D_{G}(\sigma,t)$ is \emph{not} proportional to the viscosity
$\nu,$ as one would naively expect, but to $\ln\nu.$ In Fig. 1 we plot the
dimensionless quantities $\left(  4\pi\nu\sigma_{\max}\right)  ^{-1}%
D_{G}(\sigma,t)G_{G}(\sigma,t)$ and $\Gamma_{1}\left(  4\pi\nu\sigma_{\max
}^{3}\right)  ^{-1}D_{G}(\sigma,t)$ as functions of $\sigma/\sigma_{\max}$.%

\begin{figure}
[ptb]
\begin{center}
\includegraphics[
height=1.8265in,
width=2.6792in
]%
{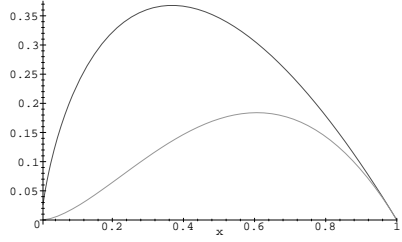}%
\caption{Plot of the dimensionless quantities (upper curve) %
 $(4\pi\nu\sigma_{\max})^{-1}D_{G}(\sigma,t)G_{G}(\sigma,t)$ %
 and (lower curve) %
$\Gamma_{1}(4\pi\nu\sigma_{\max}^{3})^{-1}D_{G}(\sigma,t)$, %
both as function of %
$x\equiv\sigma/\sigma_{\max}$ %
in the case of a Gaussian monopole.}%
\end{center}
\end{figure}
Fig.1. Plot of the dimensionless quantities (full curve) $(4\pi\nu\sigma
_{\max})^{-1}D_{G}(\sigma,t)G_{G}(\sigma,t)$ and (dotted curve) $\Gamma
_{1}(4\pi\nu\sigma_{\max}^{3})^{-1}D_{G}(\sigma,t)$, both as function of
$x\equiv\sigma/\sigma_{\max}$ in the case of a Gaussian monopole.

\subsection{Self-similar decay\label{self-similar}}

In the case of a collection of vortices that evolves self-similarly in time,
it is convenient to introduce the dimensionless independent variable
$\xi:=(\sigma-\bar{\omega})t$ and the dimensionless functions\footnote{We
assume that $A$ is finite and that $t>0.$} $\tilde{G}(\xi):=G(\sigma
-\bar{\omega},t)/At$ and $\tilde{D}(\xi):=D(\sigma-\bar{\omega},t)/(\sigma
-\bar{\omega})^{3}$. When the boundary source term $s(\sigma,t)$ vanishes, as
it does in the case of a doubly periodic domain or when the vorticity support
remains well separated from the boundary, the self-similar form of
(\ref{dG/dt}) is
\[
\frac{d\;}{d\xi}\left(  \xi\tilde{G}(\xi)\right)  =-\frac{d^{2}\;}{d\xi^{2}%
}\left(  \xi^{3}\tilde{D}(\xi)\tilde{G}(\xi)\right)  ,
\]
or
\[
\xi\tilde{G}(\xi)=-\frac{d\;}{d\xi}\left(  \xi^{3}\tilde{D}(\xi)\tilde{G}%
(\xi)\right)  +cte.
\]
Assuming that there are no singularities in the vorticity field, one has
$G(\sigma,t)\underset{\left|  \sigma\right|  \rightarrow\infty}{\rightarrow}0$
and it follows then that the constant in the last expression must be zero.
Measuring the self-similar density $\tilde{G}(\xi),$ one can solve the last
equation for the corresponding diffusion coefficient $\tilde{D}(\xi)$ and get
\[
\tilde{D}(\xi)=-\frac{1}{\xi^{3}\tilde{G}(\xi)}\int_{b}^{\xi}ds\,s\tilde
{G}(s),
\]
where the value of the lower limit of integration $b$ must be chosen according
to an appropriate ``boundary condition'' as illustrated below. If for large
$\left|  \xi\right|  $ the dimensionless vorticity density $\tilde{G}(\xi)$
decays algebraically like $\left|  \xi\right|  ^{-2\alpha}$ (see below) and
$2\alpha<3,$ then it follows that the most general decay of $\tilde{D}(\xi)$
is of the form $\left|  \xi\right|  ^{-1}+a\left|  \xi\right|  ^{(2\alpha-3)}$
with $a$ an appropriate constant.

We apply these results to two specific cases: 1) If the self-similar vorticity
distribution happens to be Gaussian, i.e., if $\tilde{G}(\xi)=\left(  \xi
_{o}\sqrt{2\pi}\right)  ^{-1}\exp\left(  -\xi^{2}/2\xi_{o}^{2}\right)  .$
Then, with $\tilde{D}(\pm\infty)=0$ as ``boundary condition'' one obtains that
$\tilde{D}(\xi)=\xi_{o}^{2}\xi^{-3}$ or, going back to the original
quantities, $G(\sigma,t)=At\left(  \xi_{o}\sqrt{2\pi}\right)  ^{-1}\exp\left[
-t^{2}(\sigma-\bar{\omega})^{2}/2\xi_{o}^{2}\right]  $ and the negative
diffusion coefficient is $-D(\sigma,t)=-\xi_{o}^{2}t^{-3}.$ This is the only
self-similar case with a $\sigma$-independent $D\left(  \sigma,t\right)  .$ In
this case, the time-evolution equation (\ref{dG/dt}) takes a particularly
simple form, namely
\begin{align*}
\frac{\partial G(\sigma,t)}{\partial t} &  =-\xi_{o}^{2}t^{-3}\frac
{\partial^{2}G(\sigma,t)}{\partial\sigma^{2}},\\
\mathrm{i.e.}\;\frac{\partial G(\sigma,\tau)}{\partial\tau} &  =+\frac{1}%
{2}\frac{\partial^{2}G(\sigma,\tau)}{\partial\sigma^{2}}\;\mathrm{with}%
\;\tau:=\xi_{o}^{2}t^{-2}.
\end{align*}
In agreement with our findings in IIA, we see that in this case the squared
width of the Gaussian distribution \emph{decreases} in time like $\tau=\xi
_{o}^{2}t^{-2}.$\newline 2) The second application is to the self-similar
distributions found by Bartello and Warn [\cite{Warn}] in their simulations
performed in a doubly periodic domain of size $A$. Qualitatively speaking,
their results can be summarized by the following expression
\begin{align}
\tilde{G}_{s}(\xi) &  =c\left(  \xi_{o}^{2}+\xi^{2}\right)  ^{-\alpha
},\;\left|  \xi\right|  \leq\xi_{M}\label{refer}\\
\tilde{G}_{s}(\xi) &  =0,\;\left|  \xi\right|  >\xi_{M}.\nonumber
\end{align}
with $\xi_{o}\simeq10,$ $\alpha\simeq0.7$ and $\xi_{M}$ growing\footnote{This
time-dependence destroys the exact self-similarity of these solutions.}
approximately like $\sqrt{t}$, from 200 to 500. The value of $c$ being such
that $\int\!d\xi\,\tilde{G}(\xi)=1.$ Vorticity values such that $\left|
\sigma\right|  t<\xi_{o}$ are associated mainly with thin filaments in the
background ``sea'' while those such that $\left|  \sigma\right|  t>\xi_{o}$
correspond to the localized vortices. At the positions with the largest
vorticity value the gradient $\nabla\omega$ vanishes, therefore, it is natural
to take as ``boundary condition'' $\tilde{D}_{s}(\xi_{M})=0.$ One gets then
the following effective diffusion coefficient in vorticity-space
\[
\tilde{D}_{s}(\xi)=\frac{\left(  \xi_{o}^{2}+\xi^{2}\right)  ^{\alpha}%
}{2(1-\alpha)\left|  \xi\right|  ^{3}}\left[  \left(  \xi_{o}^{2}+\xi_{M}%
^{2}\right)  ^{\left(  1-\alpha\right)  }-\left(  \xi_{o}^{2}+\xi^{2}\right)
^{(1-\alpha)}\right]  ,\;\left|  \xi\right|  \leq\xi_{M}(t).
\]
Going back to the original variables, this effective diffusion coefficient
reads,
\[
D_{s}(\sigma,t)=\frac{\left(  \xi_{o}^{2}+\sigma^{2}t^{2}\right)  ^{\alpha}%
}{2(1-\alpha)t^{3}}\left[  \left(  \xi_{o}^{2}+\xi_{M}^{2}\right)  ^{\left(
1-\alpha\right)  }-\left(  \xi_{o}^{2}+\sigma^{2}t^{2}\right)  ^{(1-\alpha
)}\right]  ,\quad\left|  \sigma\right|  t\leq\xi_{M}(t).
\]
The average of $\left|  \nabla\omega\right|  ^{2}$ in the thin filaments is
not zero so that at $\sigma=0$ we have
\[
D_{s}(0,t)=\frac{\xi_{o}^{2}}{2(1-\alpha)t^{3}}\left[  \left(  1+\frac{\xi
_{M}^{2}}{\xi_{o}^{2}}\right)  ^{\left(  1-\alpha\right)  }-1\right]
\simeq\frac{\xi_{o}^{2\alpha}\xi_{M}^{2\left(  1-\alpha\right)  }}%
{2(1-\alpha)t^{3}}.
\]
The origin, $\sigma=0$ is a local minimum of $D_{s}(\sigma,t),$ moreover there
is one maximum, namely
\[
\max D(\sigma,t)=\alpha^{\frac{\alpha}{1-\alpha}}\frac{\xi_{o}^{2}+\xi_{M}%
^{2}(t)}{2t^{3}}\quad at\quad\sigma^{\ast}t=\sqrt{\alpha^{\frac{\alpha
}{1-\alpha}}\left(  \xi_{o}^{2}+\xi_{M}^{2}(t)\right)  -\xi_{o}^{2}}%
\simeq\alpha^{\frac{\alpha}{2(1-\alpha)}}\xi_{M}.
\]
For large $t,$ the maximum decays like $t^{-2}$ while $D(0,t)$ decays faster,
namely like $t^{-(2+\alpha)}.$ \newline In Fig.2, we plot the dimensionless
quantities $t^{3}D_{s}(\sigma,t)\xi_{M}^{-2}$ and $4\left(  cA\xi
_{M}^{2(1-\alpha)}\xi_{o}^{2\alpha}\right)  ^{-1}t^{2}D_{s}(\sigma
,t)G_{s}(\sigma,t)$ as functions of $x\equiv\sigma t/\xi_{M}$.%

\begin{figure}
[ptb]
\begin{center}
\includegraphics[
height=2.3886in,
width=3.5034in
]%
{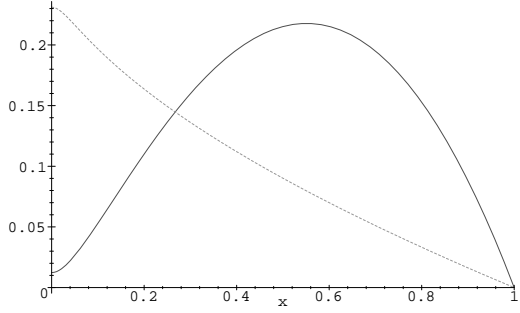}%
\caption{Plot of the dimensionless quantities (full curve) $t^{3}D_{s}%
(\sigma,t)\xi_{M}^{-2}$ and (dotted curve) $4\left(  cA\xi_{M}^{2(1-\alpha
)}\xi_{o}^{2\alpha}\right)  ^{-1}t^{2}D_{s}(\sigma,t)G_{s}(\sigma,t)$ as
functions of $x\equiv\sigma t/\xi_{M}$ in the case of self-similar decay with
$\xi_{o}=10$ and for $\xi_{M}=300$, as discussed in the main text.}%
\end{center}
\end{figure}

\section{The changes in the vorticity-area
distribution\label{vorti-redistribu}}

\subsection{General considerations\label{general vorti}}

In comparison to the case of a passive scalar, the \emph{spatial }distribution
of vorticity does not play such a central role as one realizes from Arnold's
observation that the stationary solutions of the Euler equations in two
dimensions correspond to energy extrema under the constraint of fixed
vorticity areas [\cite{Arnold1} ,\cite{Arnold2}]. Arnold's observation says
then that the stationary solutions of the 2D Euler equations, $\omega
_{S}(x,y),$ are the states with extremal values of the energy compatible with
the vorticity-area density
\[
G_{S}(\sigma):=\int\!dxdy\,\delta(\sigma-\omega_{S}(x,y)).
\]
Therefore, the vorticity-area density $G_{S}(\sigma)$ of a QSS (and the
geometry of the domain) determine $\omega_{S}(x,y),$ the spatial distribution
of vorticity in the coherent structure. From this we conclude that when
studying the process leading to the QSS in viscous fluids, special attention
should be paid to the differences between the initial vorticity-area density
$G(\sigma,0)$ and $G_{S}(\sigma),$ the one in the QSS. A convenient way of
studying these changes would be through the differences in the moments of
these distributions, which will be denoted by $\Delta_{n},$ i.e.,
\begin{align*}
\Delta_{n} &  :=\Gamma_{n}^{o}-\Gamma_{n}^{S},\;\mathrm{with}\\
\Gamma_{n}^{o} &  :=\int\!dxdy\,\omega^{n}(x,y,0)=\int\!d\sigma\,\sigma
^{n}G(\sigma,0)\;\mathrm{and}\\
\Gamma_{n}^{S} &  :=\int\!dxdy\,\omega_{S}^{n}(x,y)=\int\!d\sigma\,\sigma
^{n}G_{S}(\sigma).
\end{align*}
The dimensionless ratios $\Delta_{n}/\Gamma_{n}^{S}$ would offer a good
characterization of the changes experienced by the vorticity-area
distribution. In the next Subsections we present another possibility, linked
to the predictions of a QSS according to a statistical-mechanics approach,
which is constructed from an $\omega_{S}(\psi_{S})$ relation measured either
in experiments or in numerical simulations.

\subsection{The changes in the moments according to the statistical mechanics
approach\label{New yardsticks}}

It is proven in the Appendix that when the quasi-stationary vorticity field
$\omega_{S}(x,y)$ and the initial field $\omega(x,y,0)$ are related as
predicted by the statistical mechanical theory then the observed $\Delta_{n} $
take the values $\delta_{n}$%

\begin{align}
\delta_{n}  & =\int\!dxdy\,i_{n}(\psi),\label{recursion2}\\
\mathrm{with}\;i_{1}  & :=0\;\mathrm{and}\nonumber\\
i_{n+1}(\psi)  & =\left[  \Omega(\psi)-\frac{1}{\beta}\frac{d\;}{d\psi
}\right]  i_{n}(\psi)+\left(  -\beta\right)  ^{-1}\frac{d\Omega^{n}}{d\psi
}.\nonumber
\end{align}
defined in terms of the associted$\ \Omega(\psi)$ relation and an inverse
temperature $\beta.$ In particular, for $n\leq4,$ we have
\begin{align}
\delta_{1} &  =0,\label{chiqui}\\
\delta_{2} &  =-\beta^{-1}\int\!dxdy\,(d\Omega/d\psi),\nonumber\\
\delta_{3} &  =\beta^{-2}\int\!dxdy\left[  \frac{d^{2}\Omega}{d\psi^{2}%
}-3\beta\Omega\frac{d\Omega}{d\psi}\right]  ,\nonumber\\
\delta_{4} &  =-\beta^{-3}\int\!dxdy\,\left[  \frac{d^{3}\Omega}{d\psi^{3}%
}-4\beta\Omega\frac{d^{2}\Omega}{d\psi^{2}}+6\beta^{2}\Omega^{2}\frac{d\Omega
}{d\psi}-3\beta\left(  \frac{d\Omega}{d\psi}\right)  ^{2}\right]  .\nonumber
\end{align}
In the next Subsection we propose the use of $\delta_{n}$ as yardsticks in
order to quantify the departure of the observed changes $\Delta_{n}$ from the
theoretical predictions.

\subsection{Analysis of experimental results\label{new ExperimentAnalysis}}

In many cases, the predictions obtained from different theories are not very
different and it is not obvious which prediction agrees better with the
experimental data, see, e.g., [\cite{Majda} ,\cite{brandsX}] . Therefore, it
is important to develop objective, quantitative measures of such an agreement.

The results of the previous Subsection lead us to conclude that there is
useful information encoded in the functional dependence of $\omega_{S}%
(\psi_{S})$ and that this information can be used for the quantification of
the vorticity redistribution process in any experiment or numerical
simulation, i.e., also when (\ref{w-psi}) and (\ref{moments}) do not
necessarily hold, as long as there is no leakage and creation or destruction
of vorticity at the boundary is negligeable. We propose that one should
proceed as follows: \newline 1) Identify the predicted$\ \Omega(\psi)$
relation of the preceeding Subsection and the Appendix with the $\omega
_{S}(x,y)\approx\omega_{S}(\psi_{S})$ of the observed QSS; usually, this
satisfies $d\omega_{S}/d\psi_{S}\neq0,$ \newline 2) Determine an effective
value of $\beta$ from $\Delta_{2},$ the measured change in the second moment
of the vorticity-area distribution and from the measured $\omega_{S}(\psi
_{S})\ $relation by, confer the second line of (\ref{chiqui}),
\begin{equation}
\beta:=-\frac{\int\!dxdy\,(d\omega_{S}/d\psi_{S})}{\Delta_{2}}.\label{T2}%
\end{equation}
Since $\Delta_{2}\geq0,$ the sign of $\beta$ is always opposite to that of
$d\omega_{S}/d\psi_{S}.$\newline 3) Using this value of $\beta$ compute from
equation (\ref{recursion2}), with $\Omega(\psi)$ replaced by $\omega_{S}%
(\psi_{S}),$ the values of the yardsticks $\delta_{n}$ for $n\geq3.$%
\newline 4) The measured changes in the third and higher moments, $\Delta_{n}$
with $n\geq3,$ should be quantified by the dimensionless\ numbers $\alpha
_{n}:=\Delta_{n}/\delta_{n}.$ These numbers are all equal to 1 if and only if
the QSS agrees with the statistical mechanical prediction corresponding to the
initial distribution $G(\sigma,0)$ with equal initial and final energies,
\begin{equation}
\omega_{S}(\psi_{S})\;\mathrm{corresponds\;to\;a\;statitistical\;equilibrium}%
\leftrightarrow\alpha_{3}=\cdots=\alpha_{n}=\cdots=1.\label{agreement}%
\end{equation}
It is for this reason that one should prefer these dimensionless quantities to
other ones like, e.g., $\Delta_{n}/\Gamma_{n}^{S}$.

In closing, it may be worthwhile recalling that, at least in some cases, the
agreement between an experiment and the statistical mechanical prediction can
be greatly improved by taking as ``initial condition'' not the field at the
start of the experiment but a later one, after some preliminary mixing has
taken place but well before the QSS appears, see [ \cite{brands}]. This
improvement can be quantified by measuring the convergence of the
corresponding $\alpha_{n}=\Delta_{n}/\delta_{n}$ towards 1. In other cases, a
detailed consideration of the boundary is necessary and, sometimes, the
satistical mechanics approach may be applicable in a well-chosen subdomain,
see [\cite{chavan} ,\cite{brandsX}].

\subsection{Examples\label{examples}}

A possible $\Omega(\psi)$ relation resulting from the statistical mechanical
theory that may be compared successfully to many experimentally found curves
is
\begin{equation}
\Omega_{t}(\psi)=\Omega_{o}\frac{\sinh\chi\psi}{B+C\cosh\chi\psi
},\label{hyper}%
\end{equation}
with appropriately chosen constants $\Omega_{o},B,C$ and $\chi.$ In
particular, the flattening in the scatter plots observed in [\cite{rasmussen}]
can be fitted by this expression while the case $C=0,$ corresponds to the
identical point-vortices model [\cite{montPFA} ,\cite{montPFA2}] and the case
$\chi\rightarrow0$ with $\chi\Omega_{o}/(B+C)\rightarrow$finite, corresponds
to a linear scatter-plot. In all these cases, the $\delta_{n}$ can be derived,
as explained in the Appendix, by means of a cumulant generating function,
confer (\ref{integral2}),
\begin{align*}
\kappa_{t}(\lambda,\psi) &  =-\beta\Omega_{o}\int^{\lambda}\frac{\sinh
\chi\left(  \psi+\xi\right)  }{B+C\cosh\chi\left(  \psi+\xi\right)  }d\xi\\
&  =-\frac{\beta\Omega_{o}}{\chi C}\ln\frac{B+C\cosh\chi\left(  \psi
+\lambda\right)  }{B+C\cosh\chi\psi}.
\end{align*}
Expanding this in powers of $\lambda$ leads to the cumulants of the
microscopic vorticity distribution. In particular,
\begin{align*}
\left\langle \sigma^{2}\right\rangle _{t}-\Omega_{t}^{2} &  =-\beta^{-1}%
\frac{d\Omega_{t}}{d\psi}\\
&  =-\frac{\chi\Omega_{o}}{\beta}\frac{B\cosh\chi\psi+C}{\left(  B+C\cosh
\chi\psi\right)  ^{2}}.
\end{align*}
Integrating this over the area, one obtains that
\[
\Delta_{2}^{t}=-\frac{\chi\Omega_{o}}{\beta}\int\!dxdy\,\frac{B\cosh\chi
\psi+C}{\left(  B+C\cosh\chi\psi\right)  ^{2}}.
\]
Knowledge of $\Delta_{2}^{t}$ and of $\Omega_{t}(\psi)$ allows us to determine
the value of $\beta.$ Once $\beta$ has been fixed, one can apply
(\ref{recursion2}) and (\ref{agreement}) in order to quantify the higher-order
moments $\Delta_{n}$ and in so doing to estimate the validity of equation
(\ref{hyper}) as the prediction from the statistical-mechanics approach.

If the scatter-plot is linear, $\Omega(\psi)=k\psi,$ then (\ref{integral2})
tells us that
\[
\kappa_{linear}(\lambda,\psi)=-\beta k\left[  \lambda\psi+\frac{1}{2}%
\lambda^{2}\right]  .
\]
This implies a simple relation between the initial vorticity-area distribution
$G(\sigma,0)$ and $G_{S}(\sigma),$ the distribution in the QSS, namely
\[
\int\!d\sigma\,\exp(-\lambda\beta\sigma)G\left(  \sigma,0\right)  =\exp
(\frac{1}{2}\lambda^{2}\beta k)\int\!d\sigma\,\exp(-\lambda\beta\sigma
)G_{S}\left(  \sigma\right)  .
\]
Recall that, as already indicated immediately after (\ref{T2}), $\beta
k\leq0.$ Assuming that the Laplace transformation can be inverted, we get for
the linear case $\Omega(\psi)=k\psi,$ that
\[
G\left(  \sigma,0\right)  =\int\!d\tau\,\exp(-\frac{\left(  \sigma
-\tau\right)  ^{2}}{2\left|  \beta k\right|  })G_{S}\left(  \tau\right)
\]
In particular, if the QSS has a Gaussian vorticity-area density with a
variance equal to $\Sigma^{2}$ and the predictions of the statistical
mechanical approach hold, then the initial distribution $G\left(
\sigma,0\right)  $ must also be a Gaussian with variance equal to $\left(
\Sigma^{2}-\beta k\right)  \geq\Sigma^{2},$ in agreement with the results in [
\cite{millerPRA} ,\cite{thesis}]. Notice that knowing $G(\sigma,0)$ and the
energy of the initial state is not enough in order to determine the spatial
dependence of the initial vorticity field $\omega(x,y,0)$.

\section{Conclusions\label{Conclusions}}

In this paper, the object at the center of our attention has been the
vorticity-area density $G(\sigma,t)$ and its time evolution in
two-dimensional, viscous flows. In Section \ref{Vorticity-area} we have shown
that this density evolves according to an advection-diffusion equation,
equation (\ref{dG/dt}), with a time dependent, negative diffusion coefficient.
If vorticity is destroyed or created at the domain boundaries then the
evolution equation contains also a source term. The equation is exact: it
follows from the Navier-Stokes equation with no approximations made. For the
purpose of illustration, explicit calculations have been presented for the
case of a Gaussian monopole in \ref{Vorticity-area}\ref{anti-diffusion} and
for the case of self-similar decay in \ref{Vorticity-area}\ref{self-similar}.
We think that it will be instructive to apply these ideas to the analysis of
data that is available from numerical simulations and laboratory experiments.
In fact, it should be possible to determine this effective diffusion
coefficient $D(\sigma,t)$ on the basis of such data. Then it would be of
interest to establish a quantitative classification of the QSS formation
processes, e.g., by considering the various possible behaviours of this
coefficient and, in particular, confer\ Figs.\ 1 and 2, of $G(\sigma
,t)D(\sigma,t)$. In the case of self-similar decay, one could attempt a
closure approximation in order to predict the effective diffusion coefficient
like it is done, for very similar quantities, in the theory of passive-scalar
dispersion by random velocity fields, e.g., Kraichnan's linear Ansatz
[\cite{Kraichnan}].\newline In Section \ref{vorti-redistribu} we considered
the changes in $G(\sigma,t)$ when starting from an arbitrary vorticity field
and ending at a high-Reynolds' number, quasi-stationary state characterized by
an $\omega_{S}(\psi_{S})$ relation. In \ref{vorti-redistribu}\ref{New
yardsticks} and in the Appendix, we showed how to generate from such an
$\omega$-$\psi$ plot an infinite set of moments $\delta_{n}$, confer
(\ref{recursion2}). The changes $\Delta_{n}$ in the moments of the vorticity
distribution that are observed in a numerical simulation or in the laboratory
equal these $\delta_{n}$ if and only if the initial and final distributions
are related to each other in the way predicted by the statistical mechanical
approach. Therefore, these changes in the vorticity distribution moments can
be quantified in terms of the dimensionless ratios $\Delta_{n}/\delta_{n}.$
The deviations of the ratios $\Delta_{n}/\delta_{n}$ from the value 1 as
determined on the basis of the data gives a direct way of quantifying the
validity of the statistical-mechanics approach. In \ref{vorti-redistribu}%
\ref{new ExperimentAnalysis} we discussed how to apply this to experimental
measurements provided that the leakage or creation and destruction of
vorticity at the boundaries is negligeable. Finally, in Subsection
\ref{vorti-redistribu}\ref{examples}, we considered two relevant $\omega
$-$\psi$ relations: a linear one and a much more general one, given by
equation (\ref{hyper}). Many of these ideas should be applicable also to more
realistic systems, e.g., when potential vorticity is a Lagrangian invariant.

{\large Acknowledgements}: We have benefitted from numerous discussions with
H. Brands, from P.-H. Chavanis' comments and from the comments of the
referees. The interest shown by H. Clercx, G.-J. van Heijst, S. Maassen and
J.J. Rasmussen has been most stimulating.

\section{Appendix\label{appendix}}

We prove now that if and only if the QSS happens to coincide with the
prediction of the statistical mechanical approach, then the changes in the
moments $\Delta_{n}$ take the values $\delta_{n}$ as defined in
(\ref{recursion2}). \newline Recall that in the statistical mechanical
approach one identifies $\omega_{S}(x,y),$ the vorticity of the QSS, with
$\left\langle \sigma\right\rangle :=\int\!d\sigma\,\sigma\rho(\sigma,\psi),$
the average value of the microscopic vorticity $\sigma$ with respect to a
vorticity distribution $\rho(\sigma,\psi)$ which is given by
\begin{align}
\rho(\sigma,\psi) &  :=Z^{-1}\exp\left[  -\beta\sigma\psi(x,y)+\mu
(\sigma)\right]  ,\label{w-psi}\\
\mathrm{with\;}Z(\psi) &  :=\int\!d\sigma\,\exp\left[  -\beta\sigma\psi
+\mu(\sigma)\right] \nonumber\\
\mathrm{and\;define\;}\Omega(\psi) &  :=\left\langle \sigma\right\rangle
=\int\!d\sigma\,\sigma\rho(\sigma,\psi).\nonumber
\end{align}
In (\ref{w-psi}), $\beta$ and $\mu(\sigma)$ are Lagrange multipliers such that
the energy per unit mass and the microscopic-vorticity area distribution
$g(\sigma):=\int\!dxdy\,\rho(\sigma,\psi(x,y))$ have the same values as in the
initial distribution, i.e., $g(\sigma)=G(\sigma,0)$. Consequently, the
spatially integrated moments are given by
\begin{align}
\int\!dxdy\,\left\langle \sigma^{n}\right\rangle  &  =\int\!d\sigma
\,\sigma^{n}g(\sigma)=\int\!d\sigma\,\sigma^{n}G(\sigma,0)=\Gamma_{n}%
^{o}\label{moments}\\
\mathrm{while\;}\Gamma_{n}^{S} &  =\int\!dxdy\,\Omega^{n}(\psi)=\int
\!dxdy\,\left\langle \sigma\right\rangle ^{n}.\nonumber
\end{align}
Denote by $\delta_{n}$ the predicted change in the $n$-th moment, i.e.,
\begin{equation}
\delta_{n}=\int\!dxdy\,\left[  \left\langle \sigma^{n}\right\rangle
-\left\langle \sigma\right\rangle ^{n}\right]  =:\int\!dxdy\,i_{n}%
.\label{Deltas}%
\end{equation}
To the probability distribution $\rho(\sigma,\psi),$ defined in (\ref{w-psi}),
we associate a cumulant generating function
\begin{equation}
\kappa(\lambda,\psi):=\ln\left\langle \exp\left(  -\lambda\beta\sigma\right)
\right\rangle .\label{generate2}%
\end{equation}
This satisfies
\begin{equation}
\frac{\partial\kappa(\lambda,\psi)}{\partial\lambda}=-\beta\Omega(\psi
+\lambda)\,,\label{integral2}%
\end{equation}
as it can be shown by first noticing that
\[
\left\langle \exp\left(  -\lambda\beta\sigma\right)  \right\rangle
=\frac{Z(\psi+\lambda)}{Z(\psi)}%
\]
so that $\kappa(\lambda,\psi)=\ln Z(\psi+\lambda)-\ln Z(\psi)$ and then using
$\int\!d\sigma\,\sigma\rho(\sigma,\psi+\lambda)=\Omega(\psi+\lambda)$ for the
computation of $\partial\kappa/\partial\lambda$. Expanding both sides of
(\ref{integral2}) in powers of $\lambda$, it follows that $\kappa_{n}(\psi),$
the $n$-th local cumulant of $-\beta\sigma,$ is related to $\Omega(\psi)$ by
\begin{equation}
\kappa_{n}(\psi)=-\beta\frac{d^{n-1}}{d\psi^{n-1}}\Omega(\psi).\label{kappas2}%
\end{equation}
For example, for $1\leq n\leq4,$ these equalities read
\begin{align*}
\beta\left\langle \sigma\right\rangle  &  =\beta\Omega,\\
\beta^{2}\left[  \left\langle \sigma^{2}\right\rangle -\Omega^{2}\right]   &
=-\beta\frac{d\Omega}{d\psi},\\
\beta^{3}\left[  \left\langle \sigma^{3}\right\rangle -3\Omega\left\langle
\sigma^{2}\right\rangle +2\Omega^{3}\right]   &  =\beta\frac{d^{2}\Omega
}{d\psi^{2}},\\
\beta^{4}\left[  \left\langle \sigma^{4}\right\rangle -3\left\langle
\sigma^{2}\right\rangle ^{2}-4\left\langle \sigma^{3}\right\rangle
\Omega+12\left\langle \sigma^{2}\right\rangle \Omega^{2}-\Omega^{4}\right]
&  =-\beta\frac{d^{3}\Omega}{d\psi^{3}}.
\end{align*}
For our purposes, it is essential to eliminate from these equations products
like $\Omega\left\langle \sigma^{2}\right\rangle $ and $\left\langle
\sigma^{2}\right\rangle ^{2}$ because their integrals over the whole domain
cannot be related to known quantities. In fact, the differences $\delta_{n}$
can be directly expressed in terms of $\Omega(\psi)$ and its derivatives,
i.e., in terms of known quantities, as we now show. To this end, one considers
first the generating function of the local moment differences $i_{n}$, which
will be denoted by $i(\lambda,\psi).$ One has that, confer (\ref{Deltas}),
\[
i(\lambda,\psi):=\left\langle \exp\left(  -\lambda\beta\sigma\right)
\right\rangle -\exp\left(  -\lambda\beta\left\langle \sigma\right\rangle
\right)  ,
\]
This is related to the cumulants generating function $\kappa(\lambda,\psi),$
confer equation (\ref{generate2}), by
\[
\kappa(\lambda,\psi)=\ln\left[  i(\lambda,\psi)+\exp\left(  -\lambda
\beta\left\langle \sigma\right\rangle \right)  \right]  .
\]
Expanding both sides of this identity in powers of $\lambda$ and making use of
(\ref{kappas2}), one obtains the recursive expressions for the $i_{n}(\psi)$
given in equation (\ref{recursion2}); integrating these over the area $A,$ one
finally gets the results stated in \ref{vorti-redistribu}\ref{New yardsticks}.

\label{Biblio}\nopagebreak[4]

\end{document}